\documentclass[10pt]{iopart}

\usepackage[latin1]{inputenc}
\usepackage{graphicx}
\usepackage{tabularx}
\usepackage{iopams}
\usepackage{harvard}
\usepackage{epstopdf}

\usepackage{xspace} 

\newcommand{\ED}{\ensuremath{n}\xspace}                 
\newcommand{\RED}{\ensuremath{\widehat{n}}\xspace}          
\newcommand{\EDF}{\ensuremath{\nu_i}\xspace}            

\newcommand{\RAC}{\ensuremath{\widehat{\mu}}\xspace}        
\newcommand{\CS}{\ensuremath{\sigma}\xspace}            
\newcommand{\RCS}{\ensuremath{\widehat{\sigma}}\xspace}     
\newcommand{\RCSH}{\ensuremath{\widehat{\sigma}_\mathrm{h}}\xspace}  
\newcommand{\RCSL}{\ensuremath{\widehat{\sigma}_\mathrm{l}}\xspace}  

\newcommand{\SPR}{\ensuremath{\widehat{S}}\xspace}          
\newcommand{\SN}{\ensuremath{L}\xspace}                 
\newcommand{\RSN}{\ensuremath{\widehat{L}}\xspace}      

\begin{document}

\renewcommand{\harvardurl}{URL: \url}  

\title[Range prediction for tissue mixtures based on dual-energy CT]{Range prediction for tissue mixtures based on dual-energy CT}

\author{Christian M\"ohler$^{1,2}$, Patrick Wohlfahrt$^{3,4}$, Christian Richter$^{3-6}$, Steffen Greilich$^{1,2}$}

\address{$^1$ German Cancer Research Center (DKFZ), Heidelberg, Germany}
\address{$^2$ National Center for Radiation Oncology (NCRO), Heidelberg Institute for Radiation Oncology (HIRO)}
\address{$^3$ OncoRay - National Center for Radiation Research in Oncology, Faculty of Medicine and University Hospital Carl Gustav Carus, Technische Universit\"at Dresden, Helmholtz-Zentrum Dresden - Rossendorf, Dresden, Germany}
\address{$^4$ Helmholtz-Zentrum Dresden - Rossendorf, Institute of Radiooncology, Dresden, Germany}
\address{$^5$ Department of Radiation Oncology, Faculty of Medicine and University Hospital Carl Gustav Carus, Technische Universit\"at Dresden, Dresden, Germany}
\address{$^6$ German Cancer Consortium (DKTK), Dresden, Germany}

\ead{c.moehler@dkfz-heidelberg.de}
\vspace{10pt}

\begin{abstract}

The use of dual-energy CT (DECT) potentially decreases range uncertainties in proton and ion therapy treatment planning via determination of the involved physical target quantities. For eventual clinical application, the correct treatment of tissue mixtures and heterogeneities is an essential feature, as they naturally occur within a patient's CT. Here, we present how existing methods for DECT-based ion-range prediction can be modified in order to incorporate proper mixing behavior on several structural levels. Our approach is based on the factorization of the stopping-power ratio into the relative electron density and the relative stopping number. The latter is confined for tissue between about 0.95 and 1.02 at a therapeutic beam energy of 200 MeV/u and depends on the I-value. We show that convenient mixing and averaging properties arise by relating the relative stopping number to the relative cross section obtained by DECT. From this, a maximum uncertainty of the stopping-power ratio prediction below $1\%$ is suggested for arbitrary mixtures of human body tissues. 
\end{abstract}

%
\vspace{2pc}

%
\maketitle
%
%

\section{Introduction}

Accurate prediction of ion ranges in tissue is essential in order to fully exploit the potential of proton and ion-beam therapy in terms of an efficient target coverage using the sharp distal dose fall-off of the Bragg peak. A large part of the uncertainties currently associated to this prediction is due to the conversion of photon attenuation from computed tomography (CT) to ion stopping-power ratios (SPRs) \cite{Paganetti2012}. The present clinical standard for CT conversion is a one-to-one heuristic relation in the form of a Hounsfield look-up table (HLUT) \cite{Schneider1996,Jaekel2001}. The difficulties in CT-number-to-SPR conversion arise from the linking of different physical regimes of the involved particles, i.e. photons and ions, with their associated energy-loss mechanisms in matter.

Dual-energy CT (DECT) provides possible improvement with an alternative prediction of SPRs.
Scanning the sample with two X-ray spectra, well-separated in energy by choosing different tube voltages, allows for the determination of radiological properties \cite{Rutherford1976}. These can subsequently be used in a physics-based SPR prediction via the Bethe formula. The main challenge is the determination of the mean excitation energy (I-value), which enters logarithmically in the Bethe formula and has no analogue in the photon absorption regime. \citeasnoun{Huenemohr2014a} established a two-step method of DECT-based SPR prediction. First, a relative electron density and an effective atomic number image are calculated using a proprietary algorithm (\textsc{syngo.via} DE Rho/Z Maps, Siemens Healthcare, Forchheim, Germany). Secondly, an empirical relation, established by \citeasnoun{Yang2010}, is used to determine the I-value from the effective atomic number. The method was experimentally verified using a set of homogeneous tissue substitutes (electron-density calibration phantom 467, Gammex-RMI GmbH, Biebertal, Germany). A mean absolute deviation of 0.6\% was found in comparison with water-equivalent path lengths measured at a carbon-ion beam line.

After the successful experimental verification with tissue substitutes, it remains to be studied how DECT-based stopping-power prediction performs in the case of real tissue and ultimately for a patient. One of the main issues going towards a more realistic situation are mixtures, which appear in CT imaging on several structural levels: chemical compounds consisting of single elements in certain proportions; tissues that are mixtures of different molecular base components such as water, proteins or lipids; and CT voxels containing more than one type of tissue. All these types of mixtures are not necessarily accounted for correctly in the I-value calibration curve from \citeasnoun{Yang2010}, where only unmixed tabulated tissues of well-defined elemental composition are considered \cite{Woodard1986}. In particular, the calibration includes a gap in the effective atomic number between the soft and the bony tissue regions, which will be populated by CT voxels of a patient image due to volume averaging and therefore has to be dealt with \cite{Huenemohr2014b}.

Here, we present a mathematically rigorous approach for the coherent treatment of mixtures in DECT-based ion-range prediction by considering new quantities instead of the previously used calibration relating I-values to effective atomic numbers.

\section{Methods}

\subsection{Basic concepts and notation}

\subsubsection*{Photon attenuation}

The linear attenuation coefficient $\mu$ of photon interaction in a medium factorizes into the electron density \ED (referred to as $\rho_e$ in \citeasnoun{Huenemohr2014a}) and the photon absorption cross section per electron \CS. In dimensionless quantities relative to water, indicated by a hat on the variables' symbols, this equation reads
\begin{equation}\label{eq:attenuation}
\RAC = \RED \RCS \, .
\end{equation}
Via the common definition of CT numbers, $\xi = (\RAC - 1) \cdot 1000$, DECT provides two spectral-weighted relative attenuation coefficients, associated to the two different X-ray voltages used (typically 80 or 100 kVp and 140 or 150 kVp, respectively). The attenuation sum rule for chemical compounds or volumetric mixtures
\begin{equation}\label{eq:attenuation_sum_rule}
\RAC = \sum_i \RAC_i \, ,
\end{equation}
and \eref{eq:attenuation} yield the relation
\begin{equation}\label{eq:cross_section_sum_rule}
\RCS = \sum_i \EDF \RCS_i
\end{equation}
for the relative cross sections, where $\EDF$ are the electron-density fractions of the constituents defined as $\EDF = \RED_i / \RED$ with $\RED = \sum_i \RED_i$. The following properties arise from this definition:
\begin{equation}\label{eq:lambdai_conditions}
0 < \EDF < 1 \, \forall i \quad \textrm{and} \quad \sum_i \EDF = 1 \, .
\end{equation}

\subsubsection*{Ion stopping power and I-values}

Based on the Bethe formula, the ion stopping-power ratio can be written as
\begin{equation}\label{eq:SPR}
\SPR = \RED \frac{\SN(I, \beta)}{\SN(I_w, \beta)} =: \RED \RSN \, .
\end{equation}
with the beam's relativistic velocity, $\beta$, and the I-values of the medium, $I$, and of water, $I_w$. Neglecting shell, density, Barkas and Bloch corrections, we write the stopping number, \SN, as
\begin{equation}\label{eq:stopping_number}
\SN(I, \beta) = \ln{\frac{2m_e c^2 \beta^2}{1-\beta^2}} - \beta^2 - \ln{I} \, .
\end{equation}
Based on Bragg's additivity rule for stopping powers \cite{Bragg1905},
\begin{equation}
\SPR = \sum_i \SPR_i \, ,
\end{equation}
and \eref{eq:SPR}, the following holds for compounds or volumetric mixtures in analogy to \eref{eq:cross_section_sum_rule}:
\begin{equation}\label{eq:correction_factor_sum_rule}
\RSN = \sum_i \EDF \RSN{}_{i} \, .
\end{equation}

\subsection{I-value and energy dependence of the relative stopping number}

\begin{figure}
	\centering
	\includegraphics[width=0.49\textwidth]{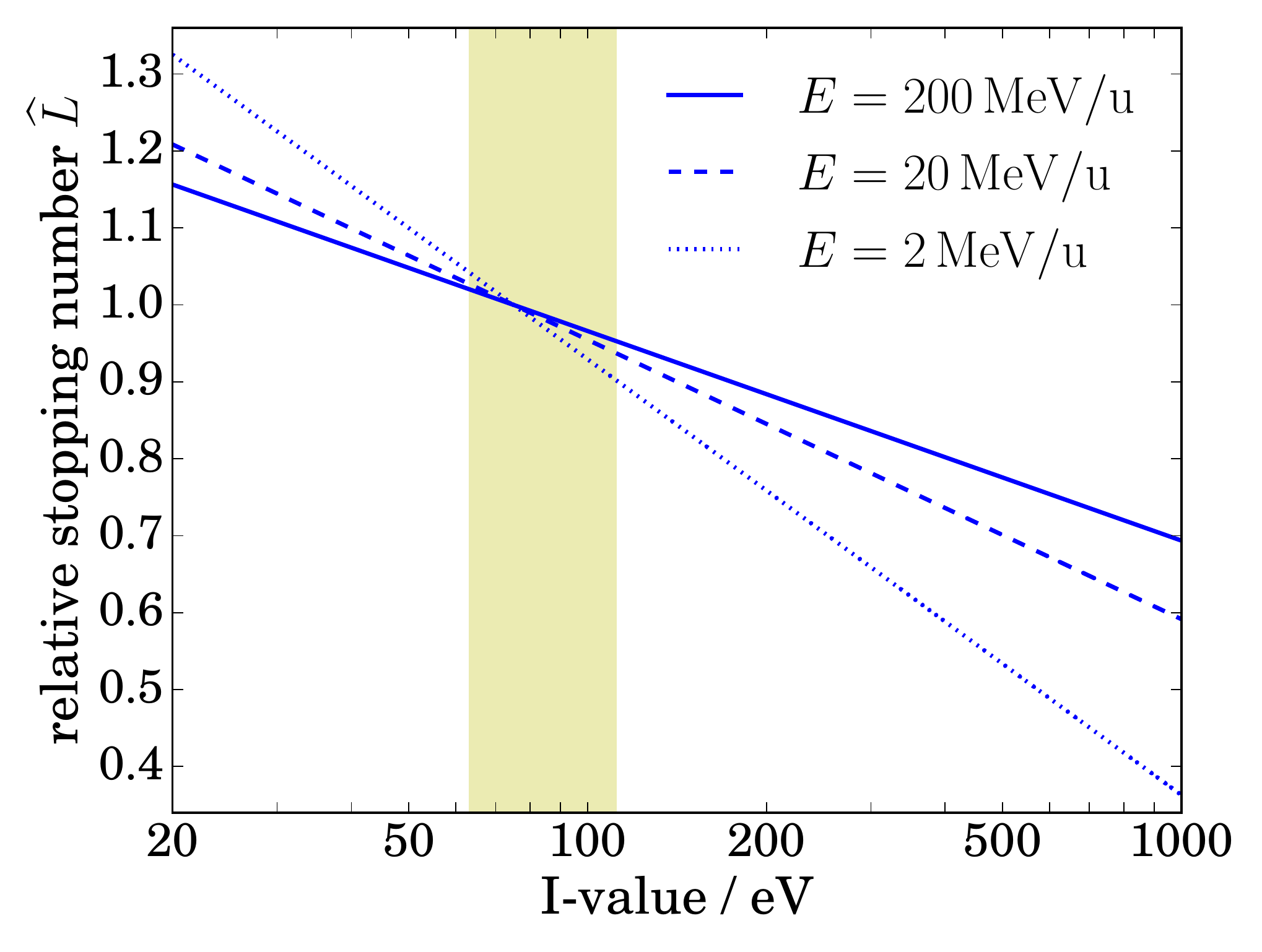}
	\includegraphics[width=0.49\textwidth]{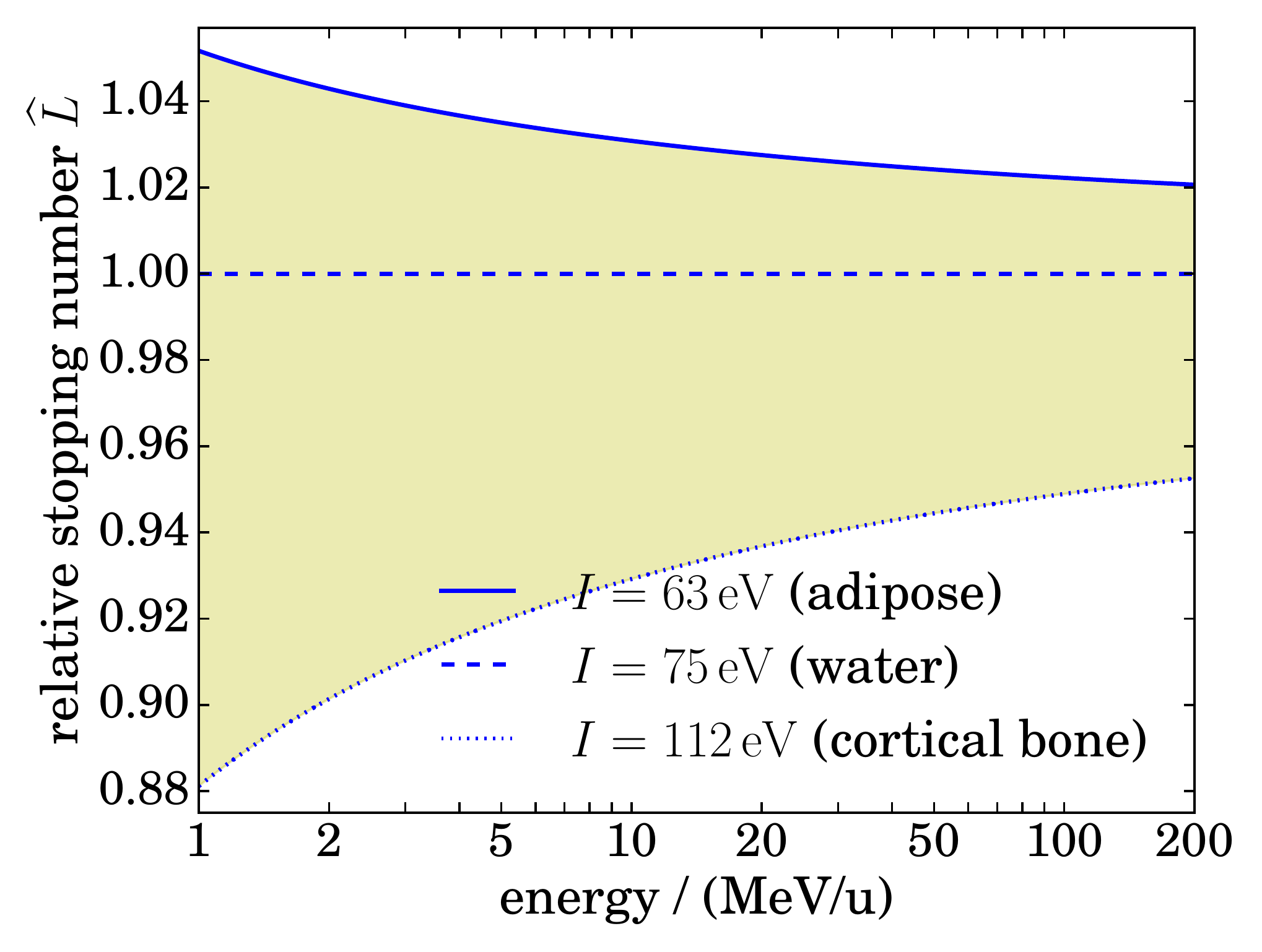}
	\caption{I-value dependence (left) and energy dependence (right) of the relative stopping number according to \eref{eq:stopping_number} with $I_w = 75$ eV. The I-value range relevant for tissue is shaded in yellow.}
	\label{fig:1}
\end{figure}

Due to the limited possible difference of $I$ from $I_w$ in human tissue and the logarithmic dependence of \RSN on $I$, \RSN is naturally bounded within a small interval around unity. For a generic therapeutic particle-beam energy of $T=200$ MeV/u, \RSN is confined between about 0.7 and 1.15 if all elements up to $Z=100$ are considered (\fref{fig:1}). In the restricted I-value range of real tissue, from about 63 eV (adipose) up to 112 eV (cortical bone), the interval shrinks down considerably to about 0.95 to 1.02, as discussed in detail below.

\Fref{fig:1} also illustrates the dependence of the relative stopping number on the beam energy, which is generally weaker than the I-value dependence in the relevant respective ranges, but gets more significant towards lower energies. It is difficult to accurately account for this effect in current treatment planning systems, as the I-value and energy-dependence cannot be easily separated with a static HLUT. This separation, however, is straightforward in DECT-based range prediction. Consequently, relative stopping numbers can be provided adapting to the decreasing energy of particles during their path in tissue. Whether this leads to a significant improvement depends on particle field configurations and the tissue traversed and therefore has to be studied with realistic clinical cases in the future.

\subsection{Prediction of the relative stopping number}

We propose to view the stopping-power ratio as the product of the relative electron density, \RED, and the relative stopping number, \RSN, according to \eref{eq:SPR}. The relative electron density can be directly determined from DECT for basically any compound or mixture in a robust method \cite{Huenemohr2014a}. As an empirical predictor variable for the relative ion stopping number, we propose to use the relative photon absorption cross section, \RCS. According to \eref{eq:attenuation}, it can be obtained by dividing the measured relative attenuation coefficient by the relative electron density, which is determined using DECT information beforehand. The relative cross section depends on the individual atomic numbers of the atoms in the considered volume and as such contains the same information as an effective atomic number, which is used as a predictor variable for the I-value determination in previous work. However, the determination of the relative cross section is more straightforward and requires less assumptions than the determination of an effective atomic number.

More importantly, the choice of the variables \RSN and \RCS enables a convenient mixing behaviour, which arises from the similarity of equations \eref{eq:cross_section_sum_rule} and \eref{eq:correction_factor_sum_rule}. Combining these, the point (\RCS, \RSN) of a composite material in the variable space of relative cross sections and relative stopping numbers can be written as a linear combination of its set of base points, $\{(\RCS, \RSN)_i\}$, according to
\begin{equation}\label{eq:convex_hull}
(\RCS, \RSN) = \sum_i{\EDF (\RCS, \RSN)_i} \, .
\end{equation}
With the conditions of \eref{eq:lambdai_conditions}, this linear combination is always convex. Thus, the part of the variable space that is filled by mixtures of a given set of base materials in all possible combinations mathematically constitutes the convex hull of those base materials.

\subsection{Construction of the (\RCS, \RSN)-space}\label{sec:construction}

Relative cross sections and stopping numbers were calculated for a number of materials for different levels of complexity from chemical elements to tabulated real tissues, using the sum rules \eref{eq:cross_section_sum_rule} and \eref{eq:correction_factor_sum_rule} respectively. The relative stopping numbers per element, $\RSN_i$, were hereby calculated via \eref{eq:stopping_number}. I-values were taken from \citeasnoun[table 2]{Seltzer1982} for single elements and from table 6 (ibid.) for elements as part of a composite material. A representative kinetic beam energy of $T=200$ MeV/u was chosen.

The relative cross sections per element with atomic number $Z$ were estimated using the parameterization
\begin{equation}\label{eq:cross_section_parameterization}
\RCS_i(Z) = a + b \cdot Z^{m}
\end{equation}
with $a=0.907 \pm 0.012$, $b=(8.5 \pm 3.5) \cdot 10^{-5}$ and $m=3.47 \pm 0.13$. The parameters $a$, $b$ and $m$ were calibrated with a data set of Gammex tissue substitutes from \citeasnoun{Huenemohr2014a}, scanned in a \textsc{Somatom Definition Flash} DECT scanner (Siemens Healthcare, Forchheim, Germany). Measured data were fitted to the equation $\RAC/\RED = a + b \sum_i \EDF Z_i^m$ using the 80 kVp CT numbers from table 2 (ibid.) and the corresponding reference data \RED, \EDF and $Z$ from table 1 (ibid.). The parameter $b$ was fixed to the water calibration point via the condition $1 = a + b\sum_i{\EDF{}_{,w} \, Z_i^m} \Leftrightarrow b=(1-a)/(0.2+0.8*8^m)$.

\subsection{Choice of the cross section}

With two measured CT numbers, DECT provides two spectral-weighted relative cross sections \RCSL (\RCSH) for the lower (higher) X-ray tube voltage. These can be used interchangeably for our purpose, as they are approximately linearly correlated. Here, we chose \RCSL, as it provides higher contrast due to the more pronounced $Z$-dependent photoelectric effect at lower photon energy. However, it is possible to use any linear combination of \RCSH and \RCSL, e.g. in the same form that is used to get the electron density from measured attenuation coefficients in \citeasnoun[eq. 14]{Huenemohr2014a} and \citeasnoun[eq. 1]{Saito2012}. The resulting quantity 
\begin{equation}
\sigma_\mathrm{mono}(E) = \alpha(E) \RCSL + (1-\alpha(E)) \RCSH
\end{equation} 
can be understood as a pseudo-monoenergetic cross section with the parameter $\alpha(E)$ determining the energy. In practical application, a particular superposition parameter (or energy $E$) could be chosen according to criteria such as the optimization of contrast-to-noise ratio or the feasibility of calibration. At this point, we would like to stress that the choice of the cross section, as well as the specific calibration procedure and the resulting parameter set in \eref{eq:cross_section_parameterization} do not affect the superposition properties of the convex hull in \eref{eq:convex_hull} and are thus without loss of generality concerning the conclusions drawn in the following.

\section{Results}

\begin{figure}
	\centering
	\includegraphics[width=0.97\textwidth]{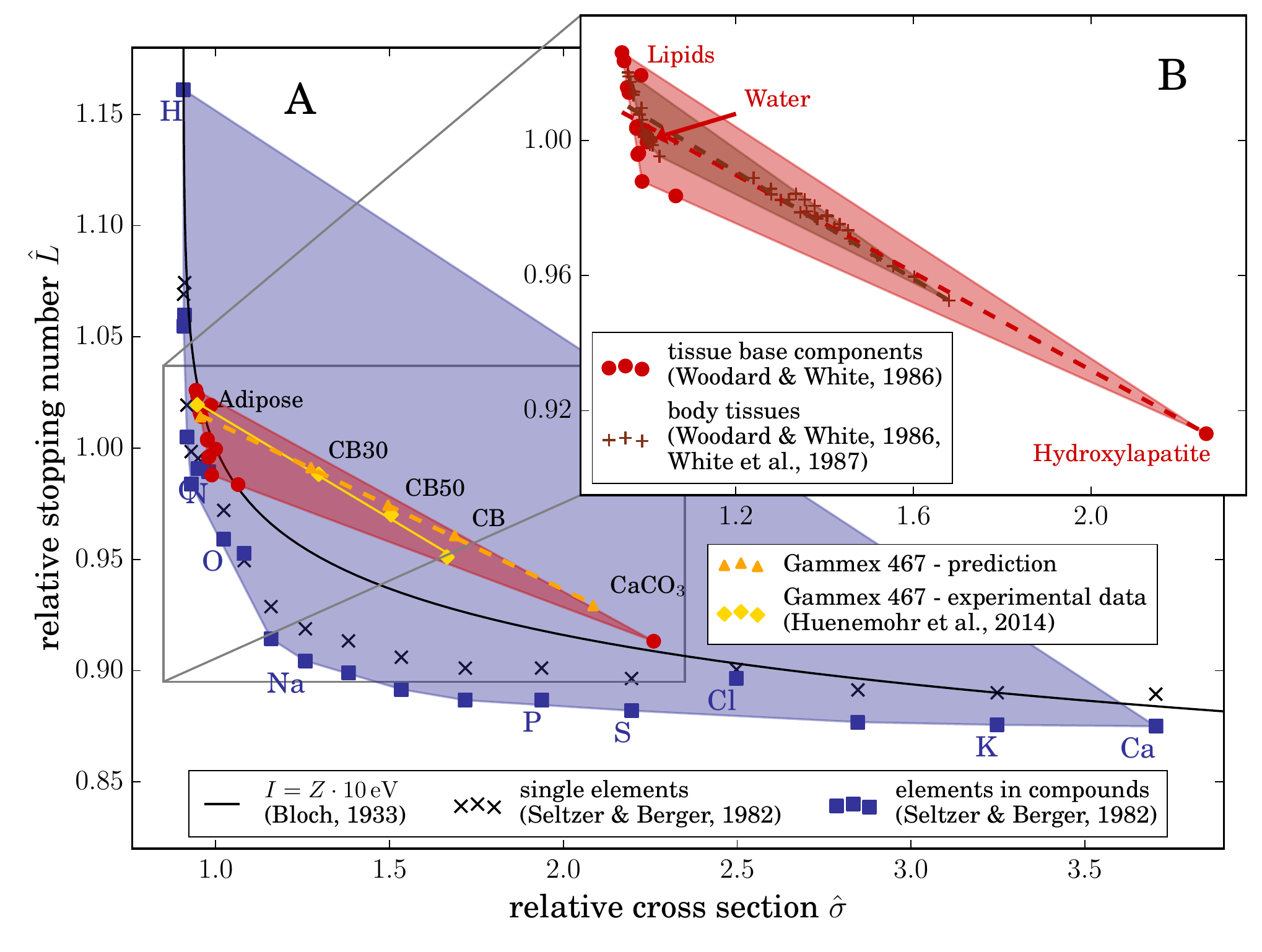}
	\caption{Illustration of the ($\RCS$, $\RSN$)-space. Single elements are shown up to $Z=20$ (Ca), with the most abundant elements in the human body annotated (panel A). Panel B zooms into the region of tissue base components and displays body tissues additionally, where the dashed lines illustrate the calibration curves described in \sref{sec:uncertainties}. The shaded areas represent the convex hull of the respective set of base points drawn in the same color. Geometrically, the convex hull is the convex polygon with the smallest area containing all the base points. For the special case of two base points, the allowed mixtures are located on the straight line connecting the two points. This linear behaviour is exemplarily shown for measurements of selected Gammex 467 tissue surrogates, along with their predictions and linear regression lines.}
	\label{fig:2}
\end{figure}

\Fref{fig:2} presents the (\RCS, \RSN)-space, constructed as described above. On the most basic structural level, i.e. for single elements, it would be straightforward to fit a unique function $\RSN(\RCS)$ and use it as a calibration curve. The simple Bloch relation $I \approx Z \cdot 10$~eV, transformed via \eref{eq:stopping_number} and \eref{eq:cross_section_parameterization}, is shown for illustration purposes. However, even in the ideal case of a perfect fit to the data points, such a calibration curve would be of very limited practical use, as a CT voxel will rarely be exclusively filled by a single element (this might nevertheless be the case for a pure one-atomic metal implant). The offset between the 'single elements' and 'elements in compounds' points reflects the factor of 1.13 that has been suggested in order to account for binding effects for all elements except H, C, N, O, F and Cl \cite{Seltzer1982}. For the latter, the values were adapted to measurements. The convex hull of the elements in compounds, as displayed in \fref{fig:2} (A), comprises the variable space that is filled admitting all possible combinations of compounds made up from elements up to calcium.

In humans, the space of possible relative stopping numbers can be drastically reduced by permitting only base components of tissue. A list of such base components with their elemental composition from \citeasnoun[table II]{Woodard1986} was augmented with hydroxylapatite, which occurs in slight variations of the chemical composition Ca$_5$(PO$_4$)$_3$(OH) as the basis of the solid structure of bones. The tabulated body tissues from \citeasnoun{Woodard1986} and \citeasnoun{White1987} are mixtures of these tissue base components. Hence, the convex hull of the body tissues is located within the convex hull of the tissue base components (\fref{fig:2}, B). The roughly linear alignment within the groups of soft and bony tissues reflects the fact that they are mixtures of two dominating components each. In the soft tissue region, these are water and lipid, whereas proteins, carbohydrates and others, being rather close to water in the (\RCS, \RSN)-plane, have a smaller influence. The tabulated bony tissues are combinations of cortical bone, which is the tissue point with maximum \RCS and minimum \RSN, and red or yellow marrow, respectively.

This linear superposition is further illustrated for a selected subset of the Gammex 467 Tissue Characterization Phantom, comprising adipose, CB30, CB50 and cortical bone (CB) as measured by \citeasnoun{Huenemohr2014a}. The elemental composition of the various CB surrogates suggests that these have been manufactured as mixtures of the material used for the adipose surrogate and calcium carbonate (CaCO$_3$). Consequently, the predictions as calculated from the elemental composition data and reference electron densities listed in table 1 (ibid.) exhibit alignment in \fref{fig:2}. Experimental (\RCS, \RSN)-points were obtained using measured CT numbers from table 2 (ibid.) and measured WEPLs and reference electron densities from table 1 (ibid.) in \eref{eq:attenuation} and \eref{eq:SPR}. A linear regression still shows high correlation ($r^2=0.9964$) compared to the predictions ($r^2=0.9996$). The deviation in slope might indicate a shift in I-values from the reference and underlines the importance of further experimental validation in the future.

\section{Discussion}\label{sec:uncertainties}

For the practical application of the proposed approach, a specific calibration has to be defined, which assigns a unique relative stopping number to the measured relative cross section in each CT voxel. Given a particular calibration curve, $\RSN(\RCS)$, the knowledge about mixing properties can then be applied to assign a calibration uncertainty, by making use of the constraints set by the convex hull of a particular base set.

Without any weighting of particular points, we assume in a first approximation a probability distribution that is uniform within the convex hull and zero outside. With this condition, the roughly triangular shape of the convex hull of tissue base components and body tissues suggests to define a central calibration line outgoing from the base point with maximum \RCS and bisecting the long upper and lower borders of the convex hull (dashed lines in \fref{fig:2}, B). The symmetric standard uncertainty of this calibration, $u_{\RSN}(\RCS)$, is accordingly defined as the difference of the upper edge and the central line divided by the square root of three \cite{JCGM2008}. The resulting linear parameterizations of the relative stopping number and its associated standard uncertainty are summarized in \tref{tab:calibration}.

\begin{table}
\caption{\label{tab:calibration}Parameters for the calibration curves $\RSN(\RCS)$ with standard uncertainties $u_{\RSN}(\RCS)$. The domain of definition, $D_{\RCS}$, corresponds to the spread in \RCS of the respective convex hull (cf. \fref{fig:2}). The linear parameterizations are of the form $X(\RCS) = a_X \, \RCS + b_X$ with $X = \{\RSN, u_{\RSN}\}$. The maximum relative uncertainty, as stated in the last column, is reached at the lower limit of $D_{\RCS}$ (soft tissue region).}
\footnotesize
\begin{tabular}{@{}ll|ll|ll|l}
\br
class & $D_{\RCS}$ & $a_{\RSN}$ & $b_{\RSN}$ & $a_{u_{\RSN}}$ & $b_{u_{\RSN}}$ & $\max\limits_{\RCS\in D_{\RCS}}{\{u_{\RSN}/\RSN\}}$ \\
\mr
body tissues           & [0.96, 1.68] & -0.0794 & 1.0861 & -0.0081 & 0.0135 & 0.6\% \\
tissue base components & [0.94, 2.26] & -0.0723 & 1.0766 & -0.0077 & 0.0175 & 1.0\% \\
\br
\end{tabular}
\end{table}
\normalsize

A maximum uncertainty of $0.6\%$ ($1.0\%$) is reached with this particular calibration considering body tissues (tissue base components) as the base materials. Combining $u_{\RSN}$ with the uncertainty of the electron-density determination from \citeasnoun{Huenemohr2014a}, $u_{\RED} = 0.4\%$, yields a maximum uncertainty of the stopping-power prediction of $0.7\%$ ($1.1\%$) for arbitrary mixtures of body tissues (tissue base components). Potential further sources of uncertainty might be found in the limited validity of Bragg's additivity rule and the uncertainty of the calculated (\RCS, \RSN)-positions of the base points that are used for the calibration itself.

\section{Conclusion and outlook}

Heterogeneities and tissue mixtures occur naturally in voxels of patient CT images and therefore have to be considered properly in any method for DECT-based ion-range prediction suitable for clinical application. We showed how this can be achieved by relating the I-value dependent relative stopping number to the relative cross section obtained from a DECT scan. Our approach makes an unambiguous quantification of uncertainties possible by exploiting the mathematical structure of the considered (\RCS,~\RSN) variable space.

The presented results can be seamlessly complemented by experiments, performing a DECT scan of appropriate samples for electron-density determination and a range measurement at an ion-beam line. Experimental constraints are hereby relaxed to a certain degree by the electron density dropping out of the equations, allowing for more flexibility in the choice and handling of samples.

\section*{Acknowledgments}

This work was partially funded by the National Center for Radiation Research in Oncology (NCRO) and the Heidelberg Institute for Radiation Oncology (HIRO) within the project ``translation of dual-energy CT into application in particle therapy''.

\section*{References}

\bibliographystyle{jphysicsB_withTitles}  
\bibliography{./references}

\end{document}